# Magnetic-field induced melting of long-range magnetic order akin to Kitaev insulators in the metallic compound Tb$_5$Si$_3$

S. Rayaprol[1], K. K. Iyer[2], A. Hoser[3], M. Reehuis[3], A. V. Morozkin[4], V. Siruguri[1], K. Maiti[2], and E. V. Sampathkumaran[1,2,5,*]

[1]*UGC-DAE Consortium for Scientific Research, Mumbai Centre, BARC Campus, Trombay, Mumbai 400085, India*

[2]*Tata Institute of Fundamental Research, Homi Bhabha Road, Colaba, Mumbai 400005, India*

[3]*Helmholtz-Zentrum Berlin für Materialien und Energie, Berlin, D-14109, Germany*

[4]*Department of Chemistry, Moscow State University, 119992 Moscow, Russia*

[5]*Homi Bhabha Centre for Science Education, Tata Institute of Fundamental Research, V. N. Purav Marg, Mankhurd, Mumbai, 400088 India*

*E-mail: sampathev@gmail.com

**Abstract**

There have been constant efforts to find 'exotic' quantum spin-liquid (QSL) materials. Some of the transition metal insulators dominated by the direction-dependent anisotropic exchange interaction ("Kitaev model" for honeycomb network of magnetic ions) are considered to be promising cases for the same. In such Kitaev insulators, QSL is achieved from the zero-field antiferromagnetic state by the application of magnetic field, suppressing other exchange interactions responsible for magnetic order. Here, we show that the features attributable to long-range magnetic ordering of the intermetallic compound, Tb$_5$Si$_3$, ($T_N$ = 69 K), containing honey-comb network of Tb ions, are completely suppressed by a critical applied field, $H_{cr}$, in heat-capacity and magnetization data, mimicking the behavior of Kitaev physics candidates. The neutron diffraction patterns as a function of $H$ reveal that it is an incommensurate magnetic structure that gets suppressed, showing peaks arising from multiple wave vectors beyond $H_{cr}$. Increasing magnetic entropy as a function of $H$ with a peak in the magnetically ordered state is in support of some kind of magnetic disorder in a narrow field range after $H_{cr}$. Such a high-field behavior for a metallic heavy rare-earth system to our knowledge has not been reported in the past and therefore is intriguing.

## 1. Introduction

In the field of magnetism, the search for quantum spin-liquid (QSL) behavior [1] due to geometrical frustration [2, 3] is an active direction of research. The strongly spin-orbit coupled Mott insulators [4] in which the magnetic ions, with the effective spin ($j_{eff}$) ½, form the (tri-coordinated) honeycomb lattice, characterized by direction-dependent anisotropic exchange interactions, was proposed to provide a promising route to achieve QSL behavior, by Kitaev [5]. Jackeli and Khaliullin [6] brought out that this Kitaev prediction might be realized in transition metal oxides. Typical examples under discussion in the literature for this proposal are: $Na_2IrO_3$, α-$Li_2IrO_3$ [7-11], α-$RuCl_3$ [12-14], and $Na_2Co_2TeO_6$ [15]. Though these insulators order magnetically, the Kitaev spin-liquid state has been claimed to be achieved by suppressing the stronger non-Kitaev spin exchange channels, responsible for magnetic ordering, by a sufficiently strong magnetic field, $H$ [16-19]. The disappearance of magnetic ordering features in bulk measurements like magnetic susceptibility ($\chi$) and heat-capacity ($C$) at a critical field ($H_c$) has been considered as a key indicator of Kitaev physics (see, for instance, Ref. 15). It is also considered important to investigate magnetic materials beyond transition-metal systems as a function of $H$, which include rare-earths ($R$) [20-22] as well those with large spin [23-26].

Keeping the above scenario in mind, we have investigated an intermetallic compound of a heavy $R$, containing honeycomb network of $R$, viz., $Tb_5Si_3$, by magnetization ($M$), heat-capacity and neutron diffraction (ND) measurements as a function of $H$ down to 2 K. We report here that, in this compound, there is indeed a suppression of homogeneous long-range antiferromagnetic order with the application of $H$, with the bulk data mimicking the behavior of commonly known Kitaev materials mentioned above. The implications of this finding are pointed out.

This compound forming in $Mn_5Si_3$-type hexagonal structure (space group $P6_3/mcm$) is made up of Tb1-Si and Tb2 layers, as shown in figure 1.

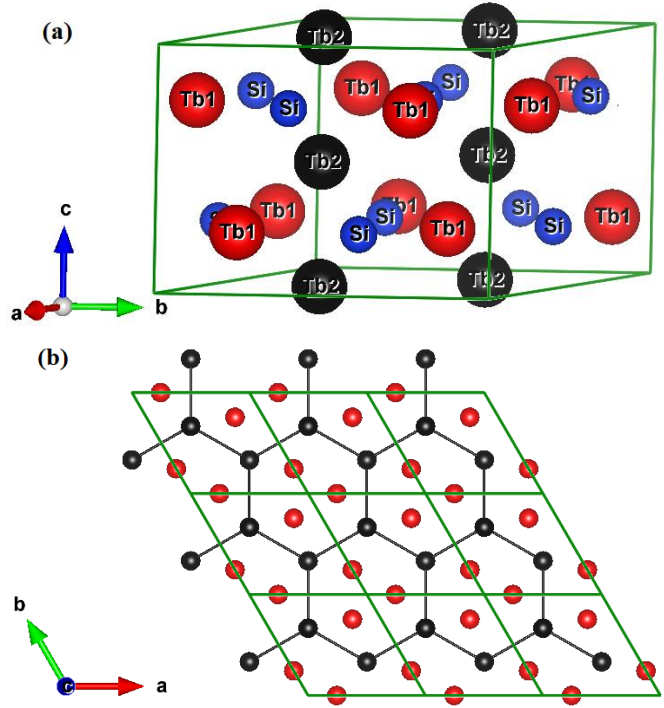

**Figure 1.** Crystal structure of $Tb_5Si_3$. (a) Unit cell, and (b) Tb2 honeycomb layer along with adjacent Tb1 layer (omitting Si). The green line represents unit-cell boundary.

Tb1 forms triangular clusters and Tb2 forms honeycomb lattice [27, 28]. Bulk measurements [29-34] suggested the onset of long-range magnetic order below $T_N$ = 69 K. The magnetic property relevant to this paper is that there is a magnetic-field-induced transition in the entire temperature ($T$) range below 69 K, as inferred from isothermal magnetization [$M(H)$] data, with the $H_{cr}$ decreasing as the temperature is increased (say, ~58 kOe at 2 K to ~20 kOe at 65 K, Ref. 29). This behavior is anisotropic [34], as it does not occur for $H$ parallel to $c$-axis, but seen only when the crystal is oriented along basal plane. There is a tendency towards saturation of magnetization well-beyond 70 kOe for the basal-plane orientation and the linear extrapolation of the high-field data to zero field yield a value of about 7.8 $\mu_B$/Tb which is less than the theoretical value ($9\mu_B$) for fully-degenerate $Tb^{3+}$ ion [34]. The intriguing property [29-34], which is the motivation for the present investigation is as follows: The magnetoresistance, MR, defined as $\{\rho(H)-\rho(0)\}/\rho(0)$, across this transition along basal plane is large, *but the sign of MR upturn is positive (instead of being negative expected for conventional metamagnetic transitions which lead to ferromagnetic alignment)*. This surprising "positive" MR jump at $H_{cr}$ is intrinsic to this compound and has been confirmed by measurements on single crystals [34]; we suggest that these come from the spin correlations characteristic of a spin-liquid, triggered by $H_{cr}$.

## 2. Experimental details

The polycrystalline sample was synthesized by arc melting together high purity constituent elements Tb (> 99.9 wt %) and Si (> 99.99 wt %) in proper proportion in an atmosphere of high purity argon gas. The samples were characterized by x-ray diffraction (Cu $K_\alpha$) to be single phase within the detection limit (< 2%) of this technique. Scanning electron microscopic image and energy dispersive x-ray analysis were further employed to ascertain the uniformity and composition of the specimen.

Heat-capacity measurements in the presence of several magnetic fields were measured by relaxation method on a piece of about 5 mg with the help of a Physical Properties Measurements System (PPMS) (Quantum Design) in the $T$-range 2 - 140 K; such measurements were reliable up to 60 kOe only, as the specimen tends to reorient at higher fields due to the torque (and hence not presented here).

Dc magnetization measurements were also performed in the range 2 - 300 K with several fields in the range 100 Oe to 90 kOe. To support the line of arguments, ac $\chi$ measurements in the $T$ region of interest were measured with different frequencies ($\upsilon$) with ac field of 1 Oe with the above-mentioned PPMS. All these measurements were performed for zero-field-cooled conditions of the specimens. Powder ND measurements were carried out on diffractometer E6 at BER II reactor of the Helmholtz Zentrum in Berlin using the neutron wavelength $\lambda = 2.43$ Å. For measuring neutron diffraction patterns, the powdered sample (~ 3g) was packed in a vanadium container and loaded on a liquid helium filled vertical field magnet with variable temperature inset. In order to avoid reorientation of grains under the application of an external magnetic fields, a mixture of deuterated methanol and deuterated ethanol was added to the sample at room temperature and cooled rapidly. On doing so, this mixture freezes around 100 K and locks the sample grains randomly, preventing any preferential orientation by magnetic field. To confirm the crystal structure of $Tb_5Si_3$, a powder pattern was collected at 300 K in the 2θ range between 7.6 - 121.5°. Neutron diffraction measurements were done in two stages. In the first study, magnetic fields up to 60 kOe were applied ($T$ = 10 K, 58 K, 72 K and 110 K) and in the second study at a later date, fields up to 120 kOe were applied ($T$ = 2 K, 5 K and 20 K). Apart from this difference in the maximum field applied, other difference was in the tighter collimation in the second study, which helped in getting better resolution in the diffraction data. However, it may be noted that both the data sets (verified at $T$ = 2K) yield similar results with respect to magnetic structure and structural parameters. The field-induced ND patterns were obtained up to 120 kOe using a shorter range between 3.4 to 73.9° at the temperatures 2, 5, 10, 20, 58, 72 and 110 K. These in-field neutron diffraction measurements were carried out at different fields, i.e., in the forward cycle of $H$ = 0, 15 kOe, 40 kOe, 60 kOe, (80 kOe, and 120 kOe in a few cases as described above) and then in the downward cycle of (80 and 60 kOe in some cases) 40, 15 and 0 kOe.

## 3. Results

*3.1 Heat capacity behavior in the presence of magnetic field:*

The results of heat-capacity measurements, measured in 0, 10, 20, 30, 40, 50 and 60 kOe, are shown in figure 2a below 85 K. Above 80 K, no significant feature is observed in the data [see figure 1 in Ref. 29]. As reported earlier, in the zero-field data, there is a sharp upturn at 69 K with decreasing temperature, followed by a peak at 67 K, familiar in order-disorder transitions, and then a gradual fall down to 2 K. This peak-temperature remains almost the same for 10 kOe, but undergoes a downward shift to 66 K and 62.6 K for an application of 20 and 30 kOe respectively. There is also a gradual suppression of the peak value. For an application of 40 kOe, the peak becomes very weak, and it appears at a lower temperature (~57 K). For 50 kOe, the peak is further smeared, appearing as a broad shoulder in the range 30 - 50 K (inset of figure 2). For 60 kOe, the peak is completely suppressed. Such a complete disappearance (around 60 kOe) of magnetic features is similar to the behavior seen in other Kitaev materials, e.g., $Na_2Co_2TeO_6$ [Ref. 15] and α-$RuCl_3$, [Ref. 16, 17]. This indicates that $Tb_5Si_3$ is transformed into a magnetically "molten" (in other words, a disordered) state around ~60 kOe. [We have also noted that the plots $C/T$ versus $T$ look similar to that of $Na_2Co_2TeO_6$, in the sense that there is a broad peak in the range 40-60 K at all fields]. The $C$ versus $T$ curves in all fields are found to overlap at low temperatures (e.g., below 40 K), thereby implying gapless behavior of the disordered magnetic state. Finally, it is worth mentioning that there is an additional weak and broad feature in the plot of $C$ versus $T$ in zero field at 70 K (marked by an arrow in figure 2a), which persists with the application of magnetic fields (but undergoing a marginal shift towards lower temperatures) and it is presumably due to the formation of antiferromagnetic clusters preceding 67 K-magnetic feature. We have also derived isothermal entropy change, $\Delta S = S(H) - S(0)$, as a function of temperature defined by the differences in the integrated area under the curves of in-field and zero-field $C/T$ plots. The values thus obtained are shown in figure 2b for a selected change of $H$.

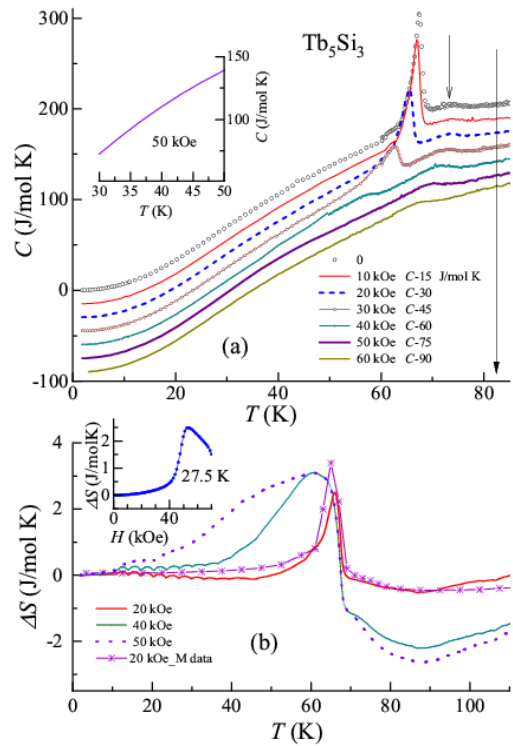

**Figure 2.** (a) Heat capacity as a function of temperature (<85 K) in the presence of various fields for $Tb_5Si_3$ to show how the peak due to the onset of magnetic order is suppressed with increasing magnetic field. The curves are shifted along y-axis for the sake of clarity, as these curves overlap well below the peaks. The small vertical arrow marks an additional weak feature at 70 K. Inset shows the measured data in the expanded form for $H = 50$ kOe in the range 30 to 50 K to show that no feature due to magnetic ordering can be seen. (b) Isothermal entropy change derived from heat-capacity data below 100 K for a change of the field from zero to $H$ to show tbe sign of the magnetic entropy in the magnetically ordered state. For comparison, the entropy change curve obtained from isothermal magnetization data (Ref. 33) for a change of field 0→20 kOe is also included; the inset shows the entropy change curve as a function $H$ for a temperature above tricritical point.

*3.2 Magnetic susceptibility behavior in the presence of magnetic field:*

Dc χ measured in different external fields are shown in figure 3 in the temperature region of interest (< 120 K) to the aim of this article.

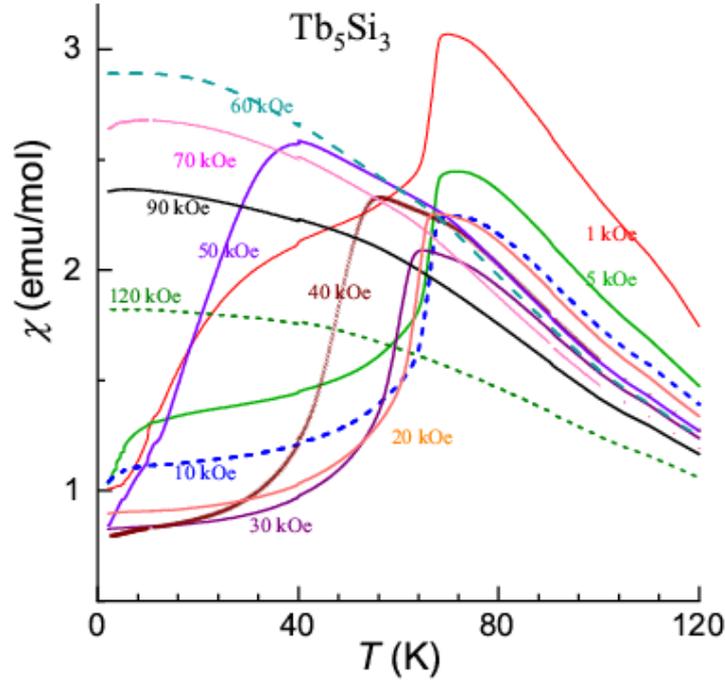

**Figure 3**. Magnetic susceptibility of Tb$_5$Si$_3$ as a function of temperature for Tb$_5$Si$_3$, measured in the presence of several fields below 120 K to show that the peak due to antiferromagnetic order is gradually suppressed with the application of magnetic field.

The χ(*T*) curves obtained in different fields do not overlap over a wide *T*-range above 70 K, deviating from the high temperature (> 220 K) Curie-Weiss region; this can be correlated to the transport anomalies in the paramagnetic state [29-34], attributed to short-ranged magnetic correlations in the high temperature magnetically disordered state. In fact, the previous zero-field ND data [35, 36] as well as the present ND studies reveal the development of weak magnetic Bragg peaks above 70 K, without any evidence for a well-defined magnetic phase transition temperature. We therefore tend to believe that short-ranged magnetic correlations indeed occurs well above 70 K, though we still use $T_N$ for 69 K transition in this paper. Returning to the magnetization behavior below 70 K, we have observed a peak in the 100 Oe curve at 69 K (not shown here for the sake of clarity of the figure), expected for the onset of antiferromagnetism, followed by a steep fall and a change of slope at about 64 K, indicating that there are subtle changes in magnetism with a lowering of temperature. The point of interest is that the features shift to the lower temperature range with increasing fields to ~68.7, ~64 K for 1 kOe, ~68.5, ~63.5 K for 5 kOe, ~67.8, ~62.7 K for 10 kOe, ~63.9, ~54 K for 30 kOe; and ~55, ~39 K for 40 kOe respectively. For 50 kOe, there is a dramatic fall in the peak temperature to ~36.8 K, but no shoulder could be resolved. The fascinating observation is that the decrease of the peak temperature for a further increase of the field does not happen at the same rate; instead, for an application of a field beyond $H_{cr}$, say, for 60, 70, 90 and 120 kOe, the peak is completely suppressed with the susceptibility tending to a constant value as temperature is lowered to 2 K. *This trend also resembles that seen in Na$_2$Co$_2$TeO$_6$ [13].* These findings, viewing together with the behavior of the peak in the *C*(*T*) described above, provide strong evidence for our conclusion that the zero-field (virgin state) long-range antiferromagnetic ordering (< 69 K) is destroyed at $H_{cr}$, mimicking quantum critical point behavior leading to spin-liquid.

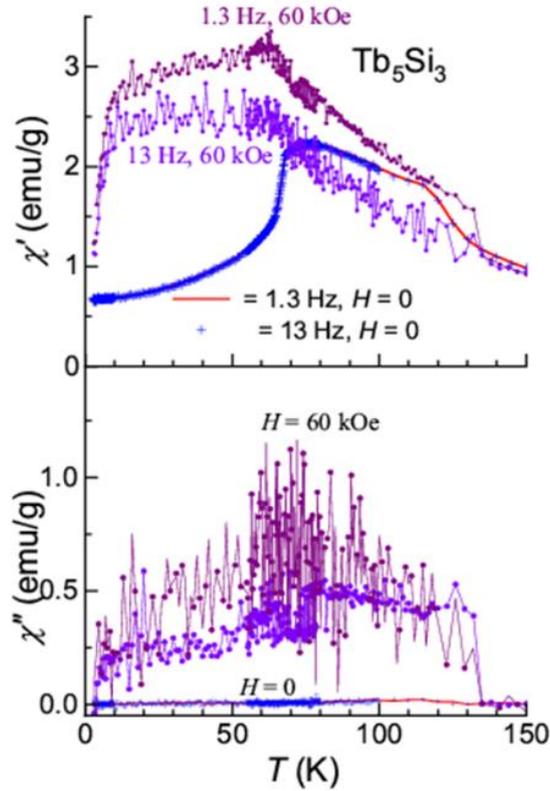

**Figure 4**. Ac susceptibility of polycrystalline Tb$_5$Si$_3$ measured in zero field as well as in 60 kOe for two frequencies. Though we show the curves for zero-field for two frequencies only, all the curves up to 1333 Hz overlap [29] and there is no signal at all in imaginary part. With respect to the data at 60 kOe, no well-defined cusp could be resolved even in imaginary part, and signal is more noisy at higher frequencies.

We now present ac susceptibility behavior. It was reported in Ref. 29 that ac χ curves measured down to 2 K in zero field overlap for all the frequencies of measurements, 1.3, 13, 133 and 1333 Hz, without showing any observable frequency dependence, thereby ruling out spin-glass behavior. Here, the ac χ data for two frequencies obtained in 60 kOe are plotted in figure 4, along with the curves for zero field. It is clear that the imaginary part (χ") does not yield any worthwhile peak, not only in zero field but also in 60 kOe, thereby ruling out spin-glass behavior, even in 60 kOe. In the presence of 60 kOe, the data also becomes noisy, becoming worse for 133 and 1333 Hz (and hence not shown for these frequencies). The drop in the real part (χ') that appears in zero field at 69 K due to magnetic ordering is suppressed in the high field data. In short, there are no spin-glass features even in high fields and the absence of any peak in the dc χ data for H > 50 kOe is also consistent with this conclusion. It may be remarked that there is a change of slope around 110 K in the zero-field ac χ data of the real part supporting short-range magnetic correlations well above $T_N$.

*3.3 Neutron diffraction results*

Neutron diffraction experiments reveal an unusual magnetic state at high fields, particularly in the close vicinity of $H_{cr}$ with an interplay between commensurate and incommensurate magnetic structures. We present here the results relevant to the aim of this article only.

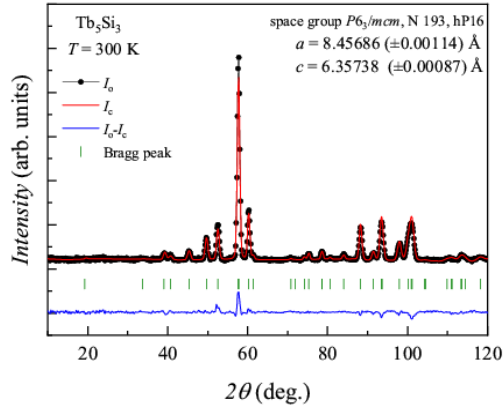

**Figure 5.** Room temperature (300 K) neutron diffraction pattern for $Tb_5Si_3$ measured on E6 diffractometer ($\lambda = 2.451$Å). The good agreement between observed ($I_o$) and calculated ($I_c$) profile confirms the phase formation of the compound. The vertical green tick marks show the Bragg peak positions as per the structural model used for Rietveld refinement. The coordinates of the Tb 6$g$ and 4$d$ atoms are ($x_{Tb}$, 0, 1/4) and (1/3, 2/3, 0), respectively. while the silicon atoms occupy the 6$g$ ($x_{Si}$, 0, 1/4) site.

In figure 5, we show the room temperature pattern along with Rietveld fitting, and this is typical of the paramagnetic state of this compound. As the temperature is lowered from 300 K to 72 K, a magnetic Bragg peak is observed (figure 6) around 5.5° in 2θ (Q ~ 0.25Å$^{-1}$), which could be indexed with a commensurate propagation vector, $k_1$ = [0, 0, ±¼]. The magnetic structure was chosen with moments given in spherical coordinates. The magnetic moments were constrained, that is, all atoms at Tb1 (at 6$g$ sites) were given same moment, and similarly all atoms at Tb2 (at 4$g$ sites) were given same moment. The ND pattern at 58 K (figure 7) is characterized by a strong magnetic Bragg peak at Q ~ 0.47Å$^{-1}$, which could be indexed with the incommensurate propagation vector, $k_2$ = [0, 0, ±0.47]. For all temperatures below this temperature, down to 2 K, the magnetic structure in zero field could be refined with this incommensurate magnetic structure only. The point relevant to the main conclusion of this paper is that ND patterns with the variation of the magnetic field up to 120 kOe at different temperatures below $T_N$ reveal the appearance of new magnetic peaks.

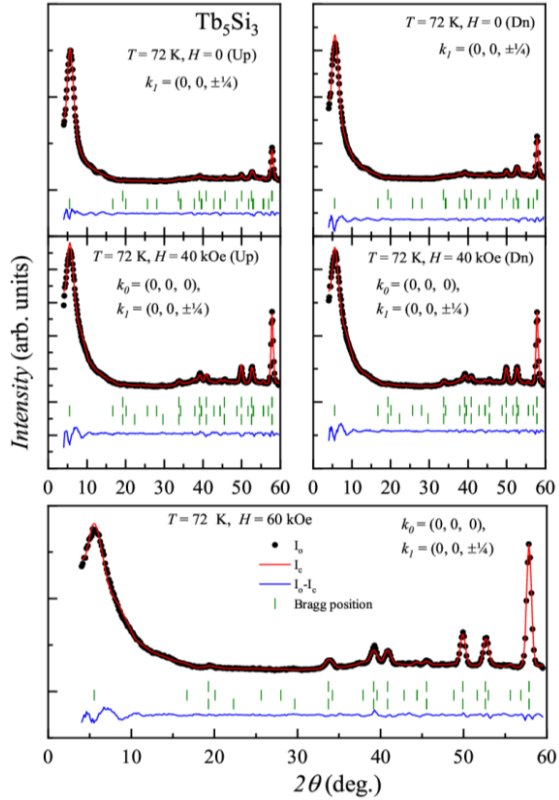

**Figure 6**. Rietveld refined ND patterns recorded at $T = 72$ K, which is just above $T_N$ in the presence of applied magnetic fields up to $H = 60$ kOe. It can be clearly seen that, for $H = 0$, a strong magnetic peak at $2\theta \sim 5°$ (Q ~ 0.248 Å$^{-1}$) is observed, which can be accounted for by choosing a magnetic structure with commensurate antiferromagnetic wave vector, $k_1 = [0, 0, ±¼]$. With increasing magnetic field a small ferromagnetic-like component with $k_0 = [0, 0, 0]$ is observed for H ≥ 40 kOe.

At 72 K, the magnetic order remains commensurate only, with the appearance of a small fraction of ferromagnetic-like pattern as $H$ increases due to the natural alignment by the field (as revealed by the increase in the intensity of nuclear peaks). However, at 58 K, though the zero-field state shows incommensurate magnetic structure, $H$ up to 40 and 60 kOe results in the formation of an additional commensurate phase ($k_1 = [0, 0, ±¼]$), apart from $k_0 = [0, 0, 0]$ propagation vector, as at 72 K, at the expense of incommensurate ($k_2 = [0, 0, ±k_z]$; $k_z \sim$ 0.47) phase. Similar observations are made in the lower temperature ND patterns with increasing $H$. That is, at all temperatures below 58 K, the external field beyond $H_{cr}$ transforms a major part of the sample into a magnetic state consisting of components of different wave vectors.

To make the main conclusion more straightforward at low temperatures, we have shown the ND patterns at various fields for 2 K in detail in figure 8a and one can clearly see the field-induced suppression of the intensity of the peak due to the incommensurate magnetic structure, indexed as "$(0,0,0)+k_2$".

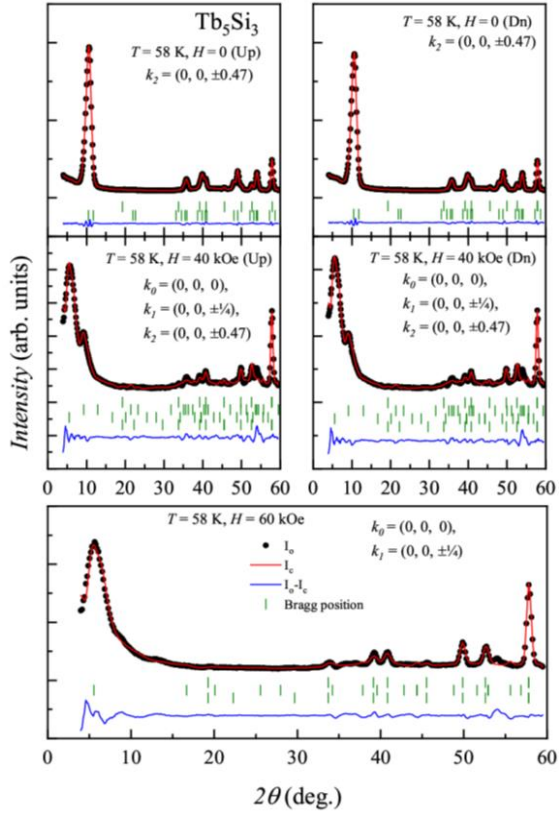

**Figure 7.** Rietveld refinement of ND patterns recorded for $T = 58$ K, which is just below $T_N \sim 69$ K, in the presence of applied magnetic fields up to $H = 60$ kOe. It can be clearly seen that, for $H = 0$, a strong magnetic peak at $2\theta \sim 10.5°$ ($Q \sim 0.467$ Å$^{-1}$) is observed, which can be accounted for by choosing a magnetic structure with incommensurate antiferromagnetic wave vector, $k_2 = [0, 0, \pm k_z]$, where $k_z \sim 0.47$. A helical magnetic structure is chosen with phase angles constrained for each Tb site. As field strength increases, there is a coexistence of multiple $k$ vectors, and for high fields ($\geq 40$ kOe), both strong antiferromagnetic peaks at $Q \sim 0.25$ and $\sim 0.47$ Å$^{-1}$ appear and, with further increase in field, the commensurate phase dominates at the expense of the incommensurate phase.

For this temperature, as $H$ is increased to 60 kOe, apart from the intensity of the antiferromagnetic peak with the wave vector ($k_2$), some peaks also show broadening. This broadening could be explained by fitting with an additional antiferromagnetic propagation vector, $k_3 = [0, ½, 0]$. On further increasing the field strength, this additional phase melts and is not recovered again while reversing the field. In figure 8b, we show the ND pattern at higher angle side in an expanded form indexing the peaks in this range. For 120 kOe (figure 8c), the profile can be fitted with a sum of the propagation vectors, two commensurate phases ($k_0$ and $k_1$) and negligible incommensurate antiferromagnetic phase $k_2$. In figure 9 we have also shown the magnetic structures corresponding to different $k$-vectors for selected temperatures / fields (refer to table 1 for those details). The field-induced phases also show a certain degree of "supercooling" effect while reversing the field for some temperatures, as in $M(H)$ curves, however with a full recovery of virgin zero field state, that is, after the field is reversed to zero from high fields (up to 120 kOe, figure 8c). The above points are presented in the tabular form in Table 1 to grasp the conclusions made above.

**4 Discussion**

From the results presented above, it is amply clear that *the features attributable to incommensurate long range magnetic ordering vanish in C(T) and χ(T) data around a certain critical field (near 60 kOe), leading to enhanced magnetic fluctuations.* However, there is no evidence for spin-glass behavior even at high fields in our ac χ measurements as mentioned above. In order to render support to the above conclusion on the evolution of magnetically molten state at $H_{cr}$, it is important to see how isothermal entropy-change vary with $H$. The overall features shown in figure 2b are in good agreement with those obtained from the magnetization data employing

Maxwells equation, as compared for 20 kOe obtained from Ref. 33. While the negative $\Delta S$ is expected above 69 K due to the suppression of magnetic disorder or short-range correlations, the sign crossover below 69 K for a conventional (antiferromagnetic to ferromagnetic) metamagnetic transition is unexpected *and the positive sign supports the onset of some form of a strongly correlated state with no long-range magnetic order, possibly a spin-liquid state - in high fields.*

Note that, as a function of $H$, there is a peak around $H_{cr}$ below $T_N$, as reproduced in the inset of figure 2b for 27.5 K from Ref. 33, clearly providing evidence for magnetic disorder around $H_{cr}$, however getting suppressed gradually at further high fields. The decrease of $\Delta S$ after the peak signals a gradual suppression of the fluctuations well beyond $H_{cr}$, which is consistent with the gradual alignment of magnetic moments by very high fields inferred from ND patterns. As mentioned at the Introduction, another strong support for the above conclusion comes from the MR data. MR versus $H$ curves are hysteretic below the tricritical point (around 20 K) and the MR curve shows an unusual increase below $H_{cr}$ while reducing the field; if this hysteresis curve is expanded by applying pressure or by doping by Lu [30, 32], there is a quadratic $H$-dependence of MR as $H$ tends to zero, followed by a linear variation of M with $H$; such a functional form is typical of dominating paramagnetic-like fluctuations. In other words, the magnetic state with fluctuations at $H_{cr}$ gets "supercooled" while reversing the field across the first-order field-induced transition [37]. As argued earlier [30], it is not possible to explain such observations in terms of a high-field ferromagnetic state alone. It is also important to note that, looking at the magnetic and magneto-transport behavior of single crystals [34], *such a field-induced disorder occurs along the basal plane only*, implying low-dimensional character of spin-liquid-like anomalies, expected for Kitaev physics also.

We now comment on the field range where the liquid-like state may be dominant. A careful look at the magnetization data of the polycrystals [see figure 2, Ref. 29] for 2 K suggests that there are indeed additional steps beyond 60 kOe, which are smeared at higher fields. The corresponding single crystal data [34] reveals a distinct step at 70 kOe (along basal plane only), in addition to the one at about 55 kOe well below $T_N$, e.g., for the 5 K $M(H)$ curve, for the orientation $H$ parallel to a basal plane; such an additional step persists even at 50 K. $M(H)$ almost saturates to 8 $\mu_B$/Tb well below $T_N$, beyond this second step, due to magnetic moment alignment expected at very high fields for a paramagnetic Tb$^{3+}$ ion. Additional intensity at nuclear reflections for 120 kOe, similar to that seen for $T > T_N$, is supportive of this line of argument. MR also exhibits an additional drop at a higher field beyond 55 kOe for the single crystals for the orientation along the basal plane (see figure 2, in Ref. 34), which is smeared for the polycrystals. *In view of these, we are tempted to propose that the spin-liquid-like state may be more prominent between two critical fields, similar to the proposal for Na$_2$Co$_2$TeO$_6$* [15].

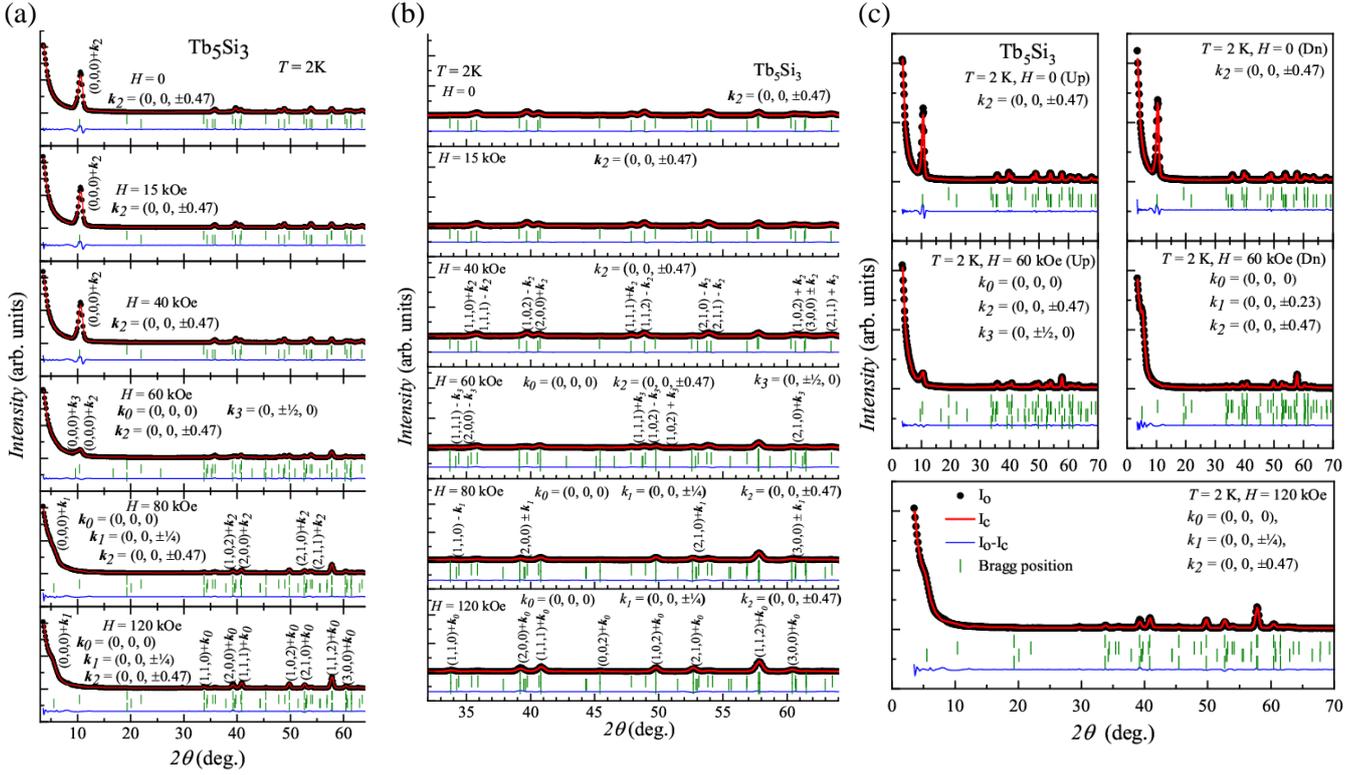

**Figure 8.** (a) Neutron diffraction patterns of Tb$_5$Si$_3$ at 2 K in the presence of different magnetic fields in the upward cycle. The data recorded at $T$ = 2 K under different magnetic fields are shown with Rietveld refinement (continuous lines through the data points) of structural and magnetic phases. Wave vectors, and difference spectra (blue lines) are also given in the figures. It is clear from this figure that the incommensurate magnetic structure is suppressed at higher fields, as inferred from the intensity of the peak around 2θ = 10º. (b) The ND diffraction patterns (upward cycle) in an expanded region in the range 32 to 64º, indexing the peaks. (c) ND patterns in downward cycle of the magnetic field for some fields.

## 5 Conclusions

We bring out that a heavy rare-earth intermetallic compound, Tb$_5$Si$_3$, containing honeycomb layers of Tb, exhibits magnetic-field-induced suppression of long-range magnetic order. Such a high-field state in the close vicinity of $H_{cr}$ is possibly a manifestation of a spin-liquid state induced by $H_{cr}$, and is also consistent with the enhancement of magnetoresistance and magnetic entropy in the positive quadrant of the respective plots as a function of field - peaking near $H_{cr}$. We bring out a parallel with the field-dependence of the bulk properties of quantum spin-liquid candidates, which turned out to fall under the category of 'promising Kitaev spin-liquids'. The fact remains that, although Kitaev physics is usually discussed in nonmetallic systems, the understanding is still incomplete as brought recently [38]. It therefore appears that this area is still in its infacncy and, in view of this, it will be of great interest to carry out further studies on this compound [39] as well as to look for such spin-liquid signatures in other honey-comb network based rare-earth intermetallic compounds. In particular, it is of interest to explore whether excitations characteristic of spin-liquids exist in inelastic neutron scattering, e.g., as in α-RuCl$_3$ [40], which is not possible in neutron diffraction studies. We hope that the knowledge gathered through such investigations could lead to a new dimension of Kitaev physics, also extending such a topic to metallic systems. A recent work by Ma [26] extending Kitaev honeycomb model to a generalized higher spin (also see Refs. 23-25] emphasizes the need to consider vaious other interactions. We therefore think that it is of interest to explore whether bond-dependent interactions arise leading to Kitaev-like physics within Ruderman-Kittel-Kasuya-Yosida interaction.

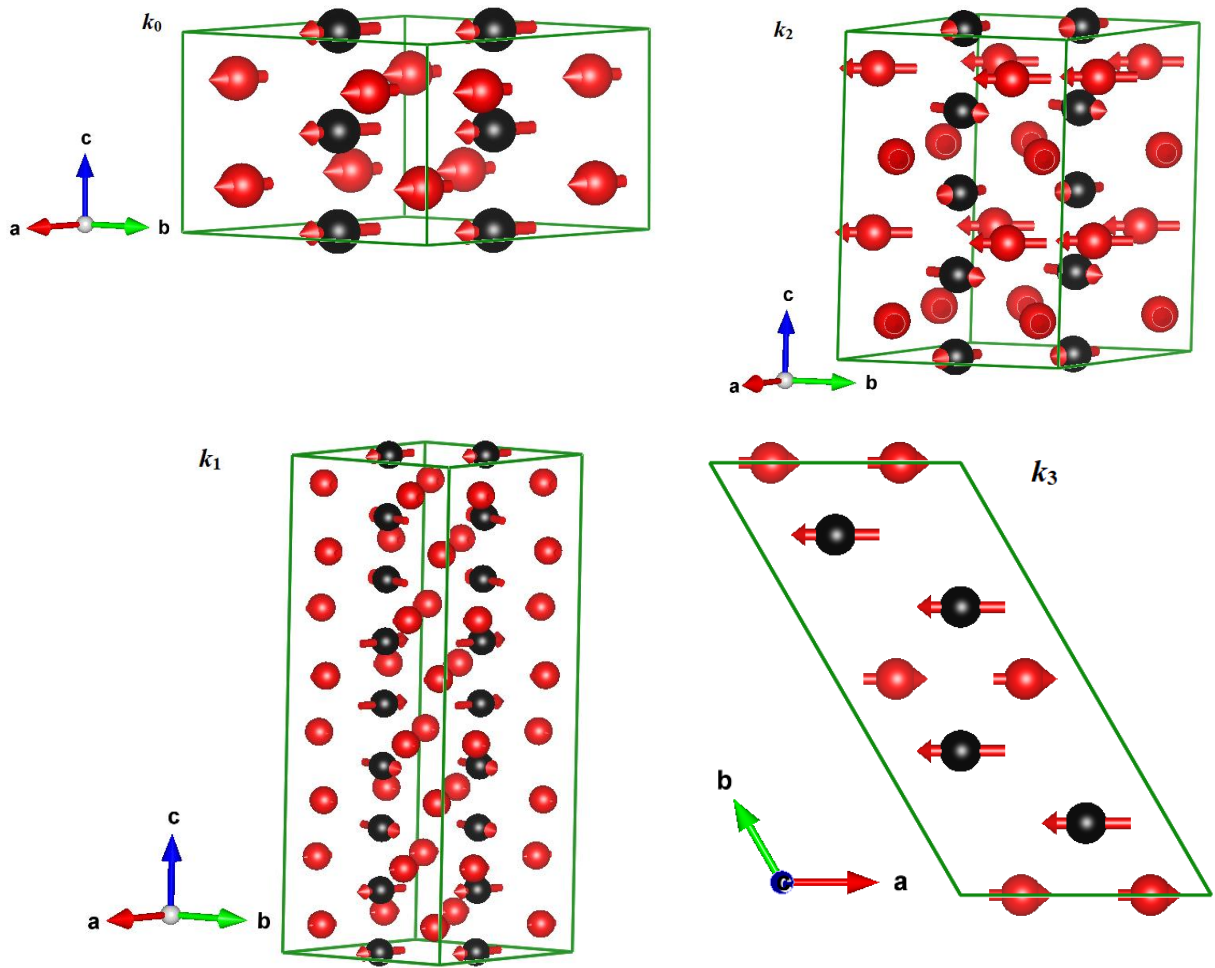

**Figure 9.** The arrangement of spins along c-axis is shown for both sites of Tb (Tb1 – Red & Tb2 – Black) for different propagation vectors required to refine the neutron diffraction profiles recorded at $T$ = 2 K but under different applied field conditions. The unit cell dimension shows the magnetic unit. All Tb ions carry magnetic moment as mentioned in the Table 1 and, for some Tb ions, the moments are pointing in such a way that the arrows are not visible for the given orientation of the figures.

**Acknowledgements**

We thank the Department of Science and Technology, Ministry of Science and Technology (Govt. of India) through sanction letters No. DST(5)/AKR/ P087/08-09/2637 and DST(5)/AKR/ P087/09-10/2169 for providing financial support to carry out ND experiments at Helmholtz-Zentrum Berlin fur Materialien und Energie, Berlin. We further acknowledge the support of Department of Atomic Energy, Government of India through Raja Ramanna Fellowship to E.V.S. K.M thanks financial support from DAE under the DAE-SRC-OI Program. E.V.S thanks Syed Hassan and G. Baskaran for an invitation to the Institute of Mathematical Sciences, Chennai, to bring this work to a completion and for fruitful discussions. The authors are grateful to Philipp Gegenwart and Deepak Dhar for fruitful discussions and critical reading of the manuscript.

**Table 1**

Results of analysis of neutron diffraction patterns of Tb$_5$Si$_3$ at 2 K for upward and downward variation of magnetic field, with details for *ab*-plane antiferromagnetic commensurate $k_1$ and incommensurate $k_2$ wave vectors (corresponding to [0, 0, ±1/4] and [0, 0, ±$k_z$] respectively, for *a*-axis antiferromagnetic commensurate $k_3$ wave vector (corresponding to [0, ½, 0]), and for *a*-axis ferromagnet ($k_0$ = [0, 0, 0]). The magnetic moments for Tb at 6*g* and 4*d* sites for each vector along with respective magnetic R-factors for magnetic phases are also tabulated; in the last column, overall profile R factors are listed.

| H (kOe) | Lattice constants (Å) | $k_0$ = (0, 0, 0) | | | $k_1$ = (0, 0, ±1/4) | | | $k_2$ = (0, 0, ±$k_z$) | | | | $k_3$ = (0, ½, 0) | | | R$_F$ (%) |
|---|---|---|---|---|---|---|---|---|---|---|---|---|---|---|---|
| | | M Tb$_{6g}$ (μ$_B$) | M Tb$_{4d}$ (μ$_B$) | R$_F$ Mag (%) | M Tb$_{6g}$ (μ$_B$) | M Tb$_{4d}$ (μ$_B$) | R$_F$ Mag (%) | M Tb$_{6g}$ (μ$_B$) | M Tb$_{4d}$ (μ$_B$) | $k_z$ | R$_F$ Mag (%) | M Tb$_{6g}$ (μ$_B$) | M Tb$_{4d}$ (μ$_B$) | R$_F$ Mag (%) | |
| 0 | a = 8.4528(14); c = 6.3534(14) | | | | | | | 8.8(2) | 8.7(2) | 0.469 | | | | 3.6 | 14.1 |
| 15 | a = 8.4535(14); c = 6.3532(14) | | | | | | | 8.7(2) | 8.7(2) | 0.469 | | | | 3.7 | 13.1 |
| 40 | a = 8.4535(12); c = 6.3535(12) | | | | | | | 8.5(2) | 8.3(2) | 0.469 | | | | 3.7 | 6.2 |
| 60 | a = 8.4515(12); c = 6.3485(15) | 3.0(2) | 4.3(4) | 7.1 | | | | 4.0(2) | 3.4(2) | 0.467 | 12.7 | 3.6(5) | 1.9(4) | 31.3 | 8.9 |
| 80 | a = 8.4449(12); c = 6.3475(13) | 4.6(3) | 5.0(4) | 4.4 | 0.3(3) | 2.9(5) | 13.1 | 0.4(2) | 0.2(2) | 0.467 | 24.9 | | | | 4.3 |
| 120 | a = 8.4375(9); c = 6.3504(10) | 8.1(9) | 7.3(9) | 6.6 | 0.6(7) | 2.3(1) | 31.6 | 0.2(3) | 0.1(2) | 0.467 | 24.6 | | | | 4.1 |
| 80 | a = 8.4412(11); c = 6.3482(12) | 6.2(6) | 6.2(6) | 6.4 | 1.2(5) | 2.7(8) | 16.5 | 0.2(2) | 0.1(1) | 0.467 | 21.3 | | | | 5.0 |
| 60 | a = 8.4453(12); c = 6.3458(13) | 4.5(3) | 4.8(4) | 5.0 | 1.2(3) [$k_z$ ~ 0.23(1)] | 3.9(5) | 10.3 | 0.2(2) | 0.1(1) | 0.467 | 20.4 | | | | 5.3 |
| 40 | a = 8.4488(13); c = 6.3585(18) | | | | | | | 4.2(1) | 4.0(1) | 0.459 | 6.9 | | | | 2.8 |
| 15 | a = 8.4530(12); c = 6.3559(11) | | | | | | | 8.6(1) | 7.7(1) | 0.463 | 3.2 | | | | 1.4 |
| 0 | a = 8.4530(12); c = 6.3562(11) | | | | | | | 8.6(1) | 7.6(2) | 0.463 | 3.1 | | | | 1.4 |